# A Point-of-Care Biosensor for Rapid Detection and Differentiation of COVID-19 Virus (SARS-CoV-2) and Influenza Virus Using Subwavelength Grating Micro-ring Resonator

Shupeng Ning,[a] Hao-Chen Chang,[b] Kang-Chieh Fan,[a] Po-yu Hsiao,[a] Chenghao Feng,[a] Devan Shoemaker,[a] and Ray T. Chen[a,b,*]

[a] *Department of Electrical and Computer Engineering, The University of Texas at Austin, Austin, Texas 78758, United State*

[b] *Omega Optics, Inc., 8500 Shoal Creek Blvd., Austin, Texas 78757, United State*

*Email：chenrt@austin.utexas.edu

## Abstract

In the context of continued spread of coronavirus disease 2019 (COVID-19) caused by SARS-CoV-2 and the emergence of new variants, the demand for rapid, accurate, and frequent detection is increasing. Besides, the new predominant strain, Omicron variant, manifests more similar clinical features to those of other common respiratory infections. The concurrent detection of multiple potential pathogens helps distinguish SARS-CoV-2 infection from other diseases with overlapping symptoms, which is significant for patients to receive tailored treatment and containing the outbreak. Here, we report a lab-on-a-chip biosensing platform for SARS-CoV-2 detection based on subwavelength grating micro-ring resonator. The sensing surface is functionalized by specific antibody against SARS-CoV-2 spike protein, which could produce redshifts of resonant peaks by antigen-antibody combination, thus achieving quantitative detection. Additionally, the sensor chip is integrated with a microfluidic chip with an anti-backflow Y-shaped structure that enables the concurrent detection of two analytes. In this study, we realized the detection and differentiation of COVID-19 and influenza A H1N1. Experimental results show that the limit of detection of our device reaches 100 fg/mL (1.31 fM) within 15 min detecting time, and cross-reactivity tests manifest the specificity of the optical diagnostic assay. Further, the integrated packaging and streamlined workflow facilitate its use for clinical applications. Thus, the biosensing platform offers a promising solution to achieve ultrasensitive, selective, multiplexed, and quantitative point-of-care detection of COVID-19.

# 1. Introduction

Since a novel coronavirus disease (COVID-19) caused by severe acute respiratory syndrome coronavirus 2 (SARS-CoV-2) was reported in late 2019, the world is continuously threatened by the potentially fatal infectious disease.[1,2] The highly contagious virus quickly spread to most continents within a few weeks and has infected more than 620 million people, including 6.5 million deaths by November 2022.[3,4] During the COVID-19 pandemic, genetic variants of SARS-COV-2 are constantly emerging and spreading as new epidemic strains.[5,6] Despite the tremendous advance in epidemiological studies and vaccine developments curbing the progress of epidemic, the variant viruses are more contagious, and may generate immune escape from innate or acquired immune responses, resulting in continued transmission around the world.[7,9] The early diagnose of suspected cases is still regarded as the best viable solution to slow down the pandemic without guaranteed preventive measures.[10]

The Omicron variant was officially named by the WHO on November 26, 2021, and quickly replaced Delta variant as the predominant strain.[11] Compared with the Alpha or Delta subvariant, Omicron has lower disease severity, hospitalization and death rates.[11-13] On the other hand, COVID-19 has become more atypical due to the mild symptoms of Omicron and thus difficult to distinguish from other infectious diseases with similar symptoms.[13,14] Among common respiratory infections, influenza presents many overlapping clinical manifestations with COVID-19, including fever, cough, sore throat, headache, fatigue, and myalgia.[15,16] However, the basic reproduction number ($R_0$) of COVID-19 (9.5 for Omicron) is much higher than influenza (0.9–2.1), which means that the SARS-CoV-2 virus is more contagious.[15,17] Hence, timely detection is significant for patients to receive tailored treatment and curbing the epidemic, considering the long-term co-existing of COVID-19 and other respiratory infectious diseases with overlapping symptoms.

In clinical practice, the primary method for COVID-19 diagnosis relies on real-time reverse transcription−polymerase chain reaction (RT-PCR). PCR-based detection shows high accuracy even in the early stages of infection, which makes it the "gold standard" in diagnosis.[18] However, RT-PCR needs advanced laboratories, expensive equipment, and medically trained personnel. Besides, RT-PCR assay requires the amplification for viral RNA, which is time-consuming and may delay the diagnosis.[19,20] Because of the increasing demand for testing and the difficulty of large-scale RT-PCR testing, reliable and rapid diagnostic methods for COVID-19 are necessary. To overcome the challenges, some lab-on-a-chip (LOC) platforms for point-of-care (POC) COVID-19 diagnosis have been developed.[9,21-23] Due to short detection time, convenient operation, and low sample requirements, these LOC techniques show great potential in clinical applications. Among these reported techniques, optical biosensors utilize the change of optical properties results from photon-matter interaction to realize the detection of analytes. Optical sensing shows several advantages in the biomedical scenario, including label-free detection, multiplexing capability, instantaneous measurements, *etc*.[10,24] In the past decades, the maturity of silicon photonics and photonic integrated circuits (PICs) technology promoted the development of optical sensors.[25] Furthermore, the compatibility with microfluidic systems offers more opportunities for LOC sensing.[26-28] Micro-ring resonators have been extensively investigated in PIC optical sensors because of their high packing density and ease of fabrication.

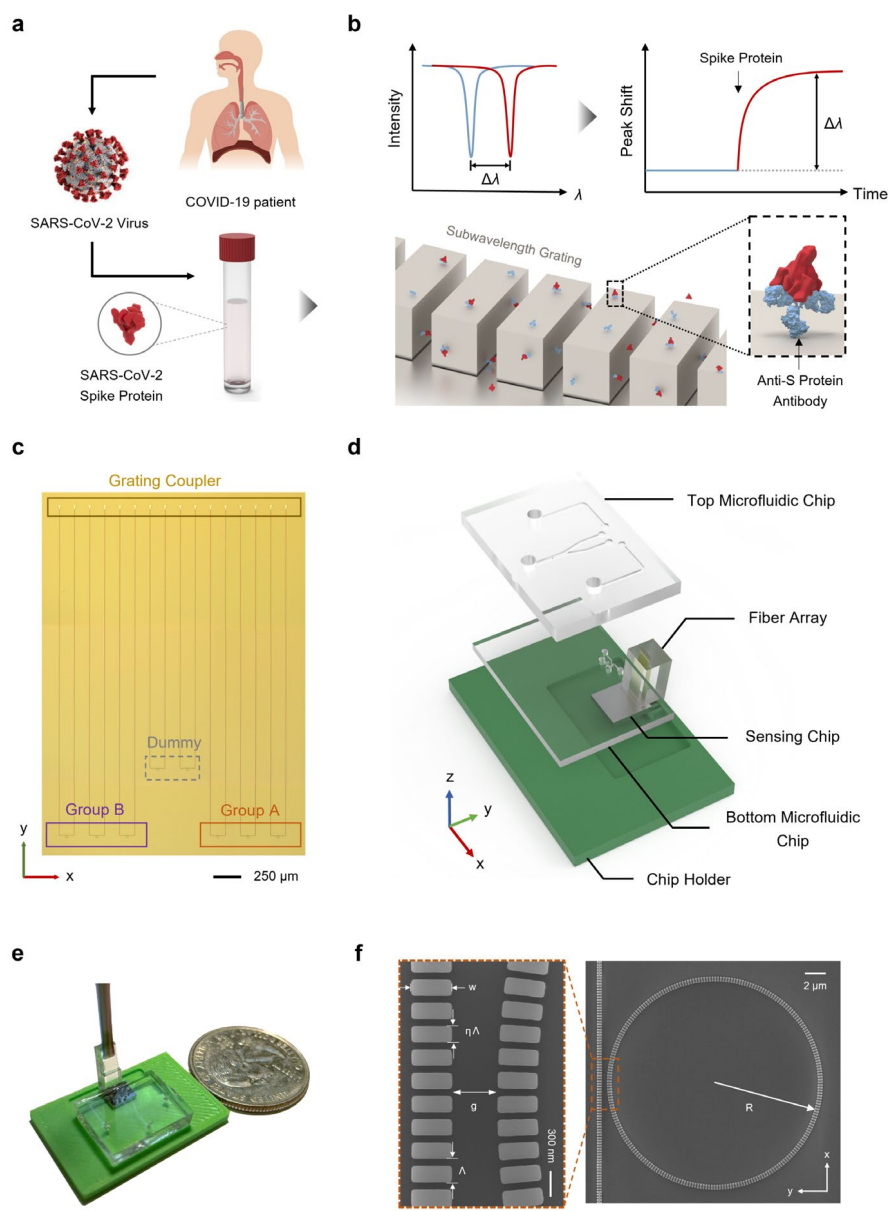

**Fig. 1. Working mechanism and design of SUMIRR-based biosensor. a** SARS-CoV-2 spike protein in phosphate buffered saline (PBS) is the target analyte for detection. **b** SARS-CoV-2 spike antibodies (blue) are conjugated on SUMIRRs as specific probes. The SUMIRR-based biosensor could quantitatively detect spike protein (red) by tracking the redshift of resonate peaks caused by antigen-antibody combination. **C** Optical micrograph of the silicon sensing chip which supports concurrent detection of two analytes. There are eight independent sensing channels, each of which has a pair of grating couplers as the input and output. Six channels are divided into two groups for two analytes, while the remaining two channels are treated as dummy group and reference. **d** An exploded view of the LOC biosensing platform. **e** Photograph of the biosensing platform with a quarter dollar for scale. **f** SEM images of the SUMIRR fabricated by E-beam lithography.

However, the limited sensitivity of micro-ring resonator impedes the application in clinical diagnostic assays that require low limit of detection (LOD).[25,29]

In this paper, we demonstrate an optical biosensing platform for the rapid detection of SARS-CoV-2 using subwavelength grating micro-ring resonator (SUMIRR) as shown in Fig. 1. Compared with conventional micro-ring resonator with strip waveguides, the subwavelength grating (SWG) structure (Fig. 1.b & f) extends the photon-matter interaction region, thus improving sensitivity.[30-33] The SUMIRRs are functionalized by SARS-CoV-2 spike antibody for the detection of SARS-CoV-2 spikes proteins (Fig. 1.b). To address the challenges posed by the untypical and diverse clinical manifestations of new epidemic strains, the LOC sensing platform enables the concurrent detection of another pathogen (influenza A H1N1 in this study) with two parallel detection groups (Fig. 1.c). To facilitate operation and improve reliability of device, we design and fabricate a double-layer microfluidic chip with an anti-backflow Y-shaped structure, which has two operating modes for surface functionalization and concurrent detection, respectively. Additionally, a three-dimensional (3D)-printed holder and specialized photonic packaging are presented to realize system-level integration. In the past few years, various diagnostic assays for COVID-19 were widely available (summarized in Table S1). However, a particularly promising solution, including packaging, testing and data processing, for POC use that can achieve concurrent quantitative detection of multiple analytes is not yet available. Our SUMIRR-based sensor detects target SARS-CoV-2 antigen with a conservative LOD of 100 fg/mL (1.31 fM). Furthermore, cross-reactivity tests for SARS-CoV-2 and influenza indicate the specificity of the optical diagnostic assay. The integrated device and auxiliary portable terminal, which offers real-time data processing and the potential to interface with electronic medical records, make the platform promising for POC diagnosis.

## 2. Device design

### 2.1 Design of the sensing platform

In considering the design of a POC platform for quantitative detection, we aimed to develop a device with clinical practicality, high accuracy and the capacity to integrate with digital systems. Toward this goal, we designed a microfluidic chip that supports dual-channel concurrent detection with a 3D-printing polylactic acid (PLA) chip holder (Fig. 1.d and e).

For concurrent detection, two parallel channels need to be placed on the silicon sensing chip with limited area (5mm × 5mm), thus the microfluidic device was designed as a double-layer structure for reliability and ease of operation (Fig. 1.d &Fig. 2.a). As shown in Fig. 2.b, the bottom microfluidic chip was designed with two independent microchannels, and the unilateral channel covers three SUMIRRs as a sensing group. Each channel had an inlet and an outlet connected to the top microfluidic chip through via holes. In addition to the left and right ports connected to the two channels in the bottom layer respectively, the top microfluidic chip has a common port connected to both channels. Besides, the common port is connected to a special Y-shape anti-counterflow splitter that can operate in two modes. In single-channel mode, the left/right port works as inlet, while the common port works as outlet. When the sample flows toward downstream (common port) from one branch of the Y-shaped structure, the significant difference in flow resistance of downstream and the other branch will "block" the other branch, thus avoiding the cross-contamination caused by counterflow (Fig. 2.c and Video S1). The other mode is the dual-channel mode, where the common port works as inlet while left port and right port are both outlets. In this mode, the structure is a splitter that bifurcates the upstream fluids towards two sensing groups (Fig. 2.d and Video S1). The anti-counterflow splitter is simulated by the finite element method to optimize chip functionality.[34-36] As shown in Fig. 2.e, the Y-shape structure can avoid counterflow in the single-channel mode without losing the bifurcating function for dual-channel mode.

For better packaging and integration, we designed a 3D-printed holder. The upper surface of holder is slotted, and the depth equals the thickness of sensing chip (0.75 mm). The slot area is larger than that of sensing chip, which aims to utilize the limited elastic deformation of PDMS to eliminate manufacturing errors from 3D printing while providing enough support for the microfluidic chip. Besides, we designed a comb structure with a height slightly less than slot depth, which serves two purposes. First, the comb structure helps to alignment; meanwhile, the fiber array is fixed on the holder by UV light adhesives, and the comb structure could provide a larger contact area resulting in higher bond strength.

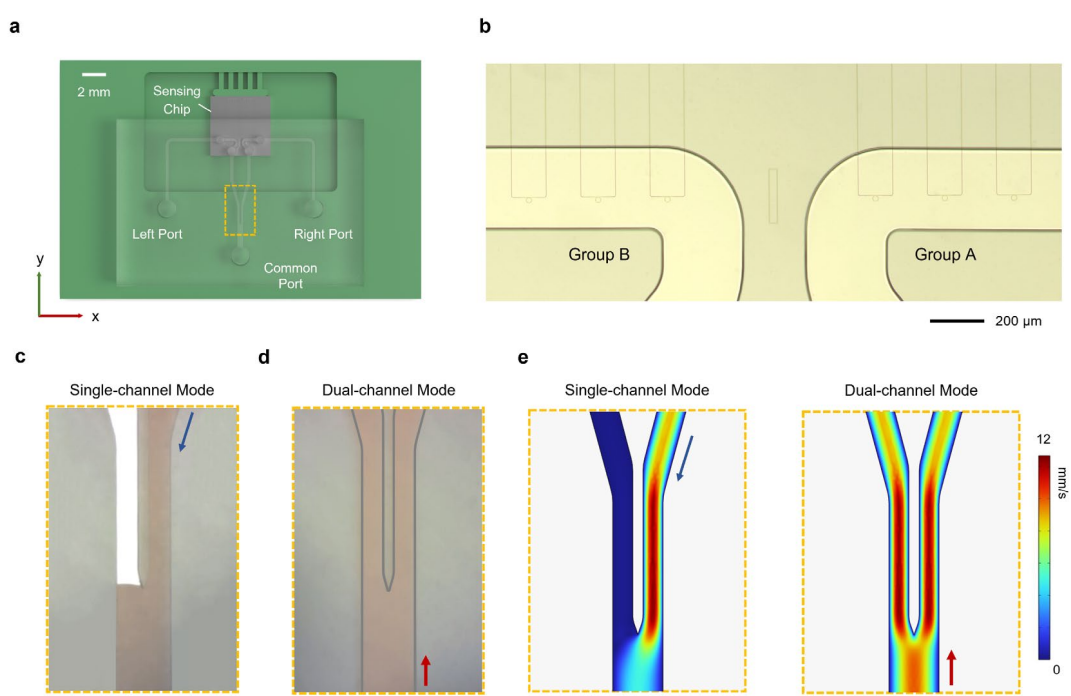

**Fig. 2. Overview of microfluidics design. a** A double-layer PDMS microfluidic chip was designed for dual-channel concurrent detection. Three ports on the top layer play different roles in different operation modes. **b** The optical micrograph illustrates that the bottom microfluidic chip bonded on the sensing chip has two independent channels, and each channel covers three resonators as one sensing group. **c & d** The working mechanisms of Y-shape anti-counterflow structure works in two different modes. Red pigment was added to DI water for demonstration purpose. **e** Hydromechanical simulation results demonstrate the velocity field distribution of two modes. In single-channel mode, the flow rate is 5 μL/min, and the right port works as inlet. While the common port is inlet in dual-channel mode with a flow rate of 10 μL/min. Details of the numerical simulation are in the SI.

## 2.2 Design of optical biosensor

The SUMIRR consists of a micro-ring and a bus waveguide constructed by periodic pillars with a period much smaller than the operating wavelength.[30,31] One advantage of the subwavelength structure is that the optical properties, including effective index, loss, guiding capabilities, *etc.*, can be modulated by geometric topological designs.[31,32] Importantly, SUMIRRs show good potential for biosensing because the periodic structure increases the effective sensing region, including not only the top and the side of pillars, which leads to more significant photon-matter interaction and higher sensitivity.[37,38]

In this study, SUMIRRs are fabricated on silicon on insulator (SOI) wafers using E-beam lithography.[32,39] Specifically, the silicon pattern sits on the $SiO_2$ buried layer (Fig. 1.f and Fig. S7), and SUMIRRs are covered by aqueous cladding as sensing units. To ensure that guided mode exists in SWG waveguides, the effective refractive index of SWG waveguide ($n_{eff}$) must be bigger than the refractive index of bottom buried layer ($n_{SiO2}$ = 1.45) and cladding layer ($n_{clad}$ ≈ 1.35).[40] The optimized design shall provide large overlap integral with analytes while maintaining decent waveguide propagation loss.[41] The resonant wavelength $\lambda_{res}$ can be expressed as:

$$\lambda_{res} = \frac{2\pi \cdot R \cdot n_{eff}}{m} \quad (1)$$

where $R$ is the radius of micro-ring and $m$ is a positive integer denoting the mode order. In the biosensing scenario, biochemical reactions, such as antigen-antibody combination and DNA hybridization,[42,43] could affect the photon-matter interaction and thus change $n_{eff}$ according to the Lorentz-Lorenz relation, which manifests in the shift of resonant peaks (Fig. 1.b).[44]

To evaluate the device's performance, we use quality factor $Q$ and bulk sensitivity $S_{bulk}$ to quantify the sensing properties of device. A higher $Q$ means that light has a longer lifetime in the resonator, and thus provides stronger interaction with analyte.[37] $Q$ is defined as Eq (2).

$$Q = \omega_{res} \frac{E}{dE/dt} = \frac{2\pi \cdot n_g}{\lambda_{res} \cdot \alpha_{[m^{-1}]}} \approx \frac{\lambda_{res}}{FWHM} \quad (2)$$

Here $\omega_{res}$ is the resonant frequency, $E$ is the mode's electric field intensity, $\alpha$ represents the total loss in the resonator and $n_g$ is the group index of the mode which can be approximated to $n_{eff}$ within a slight shift of $\lambda_{res}$.[45] From the experimental point of view, $Q$ can be approximated by the ratio of $\lambda_{res}$ to the full width at half maximum (FWHM) bandwidth of resonant peak.[46,47] A higher $Q$ is desirable because of sharper peaks which are easier to detect. The bulk sensitivity of SUMIRR $S_{bulk}$ was defined as the slope of peak shift versus change in $n_{clad}$,[48] *i.e.*,

$$S_{bulk} = \frac{\Delta\lambda_{res}}{\Delta n_{clad}} = \frac{\lambda_{res}}{n_g} \cdot \frac{\partial n_{eff}}{\partial n_{clad}} \tag{3}$$

where $S_{bulk}$ is often written as nm/Refractive Index Units (nm/RIU). Considering the influence of $Q$ and $S_{bulk}$ on sensing performance, the inherent limit of detection (ILOD) of ring resonator can be define as follows:[49]

$$\text{ILOD} = \frac{\lambda_{res}}{Q \cdot S_{bulk}} \tag{4}$$

Taking ILOD as an evaluation criteria, the optimal geometric parameters of SUMIRRs are determined by full 3D finite difference time domain method (Lumerical FDTD Solutions) and the testing after fabrication. The details of simulation and design parameters are shown in SI.

## 3. Materials and methods

### 3.1 Device fabrication and packaging

The SUMIRR-based sensing chip was custom manufactured by Applied Nanotools Inc. using E-Beam lithography on an SOI wafer with a 220 nm active layer and a 2 μm buried oxide layer (Soitec). Before use, the chip was soaked in piranha for 30 min, followed by deionized (DI) water (W4502, Sigma-Aldrich) and isopropanol (200440, CMC Materials) washing, then dried under nitrogen ($N_2$) stream.

Microfluidic microchannels were fabricated from an SU-8 photoresist (SU-8 2035, Kayaku Advanced Materials) male mold patterned on a 4-inch silicon wafer (71893-07, Electron Microscopy Sciences) using standard soft-lithography processes.[50] The height of microchannel determined by the thickness of SU-8 coating was 50 μm. Microfluidic channels were cast from the male mold with a 10:1 mixture of PDMS base and curing agent (SYLGARD 184 silicone elastomer kit, Dow) and cured at 90 °C for 40 min. The bottom and top layer thicknesses are 3mm and 6mm, respectively. Because of the small chip size, the spacing between two independent microchannels in the bottom layer should be small to ensure sufficient space for packaging. A 240 μm spacing between two 300μm-width channels showed good stability at a flow rate ≤ 100 μL/min without leakage. The via holes on bottom chip and 3 ports of top layer were punched by 0.75 mm and 1 mm punchers (PT-T983, Darwin Microfluidics), respectively.

The holder for integration was designed by CAD software (AutoCAD 2023, Autodesk) and printed by a 3D printer (FLOW, Craftbot) with PLA polymers. The printing thickness of single PLA layer was 100 μm. After cleaning with IPA and DI water, the bonding surface of sensing chip and PDMS were treated with UV rays in a UV ozone cleaner (T10X10/OES, UVOCS Inc.) for 8min and 5 min, respectively. Then, sensing chip and bottom PDMS were aligned manually according to the pre-designed alignment marks with IPA lubrication and baked in an oven at 90°C for 1 h. The bonding for bottom and top PDMS layers was the same as the above step, except that the UV treatment time was 5min. For the precise alignments of fibers and grating couplers, the packaging holder and fiber array (pitch 127um 8 degrees, Gloriole Electroptic Technology Corp) were anchored to a mechanical stage, as shown in Fig. S1. The input fiber was connected to a broadband LED (DL-BX9-CS5403A, DenseLight), and the output was connected to a C-band optical spectrum analyzer (OM-1C2MM353, Optoplex). The position of fiber array is fine-tuned to maximize the output power without losing sufficient responses from other channels. When the optimal position was determined, dropped UV light adhesives glue (37-322, Edmund Optics) on the comb structure of holder, then treated with UV light overnight for curing.

## 3.2 Surface functionalization of micro-ring resonators

The packaging device was treated by UV rays for 8 min to form hydroxyl groups (-OH) on the SUMIRR surface and remove organic contaminants. 2%[*] organosilane reagent (3-aminopropyl) triethoxysilane (APTES) diluted in 95% ethanol solution was pumped to microfluidic channels via a PTFE tubing (0.6 mm ID x 1 mm OD, Uxcell) at 5 μL/min. The flow rate was controlled by a syringe pump (70-4504, Harvard Apparatus). APTES condensed with -OH results in the formation of siloxane bonds (Si-O-Si) on UV-treated silicon surfaces.[51,52] Notably, it is the oxide layer on Si surface that participates in silanization. The following experiment and related studies confirmed that the natural silicon oxide layer is sufficient for silanization.[53-56] The unbounded APTES was removed by 95% ethanol washing at 10 μL/min for 20 min, following by drying under $N_2$ stream. Then the device was baked at 95°C for 1 h to enhance bonding stability. After silanization, an aqueous solution of 2.5% glutaraldehyde was introduced at 5 μL/min for 1 h, and then the chip was washed with PBS solution (J61196AP, Thermo Fisher Scientific) at 10 μL/min for 20 min. After silanization, an aqueous solution of 2.5% glutaraldehyde (GA) was introduced at 5 μL/min for 1 h. One aldehyde group bound to the surface expressing $-NH_2$ (from APTES) and the other aldehyde group for further crosslink with bioreceptor protein. Then the chip was washed with PBS solution at 10 μL/min for 20 min.

For antibody immobilization, 10 μg/mL antibody in PBS buffer was introduced for 40 min, followed by 20 min PBS washing. There is a potential issue with crosslinking by GA since the aldehyde groups are nonspecific to proteins. Hence, bovine serum albumin (BSA) was used to block the aldehyde sites that did not combine with antibodies to avoid nonspecific bonding and ensure the peak shift is entirely due to specific antigen-antibody combination.[57,58] 0.4 mg/ml BSA (B8667, Sigma-Aldrich) was added to block the remaining sites without antibody coating. In this work, we used two antibodies, *i.e.*, SARS-CoV-2 spike antibody (40150-D003, Sino Biological, Inc.) and pan influenza A nucleoprotein antibody (40205-R063, Sino Biological, Inc.). Notably, the immobilization for two antibodies was independent. Specifically, the SARS-

---

[*] All concentrations expressed as percentages in this paper are volume ratios.

CoV-2 antibody was introduced from one port (left port or right port), followed by BSA blocking, while the immobilization of influenza antibody was realized through the other port with the same process. The SARS-CoV-2 antibody with green fluorescence was purchased from Thermo Fisher Scientific (53-6491-82).

### 3.3 Sample preparation for antigen detection

The SARS-CoV-2 Spike S1-His recombinant protein (40591-V08H, Sino Biological, Inc.) and influenza A H1N1 nucleoprotein (40205-V08B, Sino Biological, Inc.) were diluted in PBS buffer to given concentrations depending on experimental requirements.

### 3.4 Sensing measurement

Measurements were conducted on an optical table. The fiber array was connected to a broadband LED and an OSA which communicated with PC through serial communication. The output spectrum signals were filtered to remove background noise. The real-time peak tracking and analysis were realized with custom programs (LabView, National Instruments). Plots and histograms were created and analyzed in Origin (Origin 2022, OriginLab).

Optical micrograph and videos were taken on the viewing stage of a microscope (BX51, Olympus) under 5× magnification. Fluorescent images were captured with a spinning disk confocal system (CSU-W1, Nikon) and supporting commercial software for analysis.

## 4. Results and Discussion

### 4.1 Bulk Sensitivity Analysis

Before detecting biological samples, we first measured the bulk sensitivity of fabricated SUMIRRs and tested the sensing performance after packaging. Seven samples, including DI water and six PBS solutions with different concentrations, were introduced to the device in order of concentration from low to high at 10 μL/min. The transmission spectra of SUMIRRs are shown in Fig. 3.a. The free spectral range (FSR) around 1550 nm is 14 nm, and the FWHM of resonate peaks is ~ 0.93 nm, corresponding to a $Q \approx 1650$. According to transmission spectra, the signal amplitude slightly decreases with 0.3 dB/nm as wavelength increases. Therefore, the resonate peaks near 1535 nm were selected as analytic targets for higher signal to noise ratio (SNR). In Fig. 3.b, we took the resonate wavelength with DI water cladding as a baseline and tracked peak shifts in real time. The relationship between peak shift and refractive index change was drawn and fitted linearly in Fig. 3.c. The result shows that a linear function fits well with a regression correlation coefficient $R^2 = 0.999$, and the slope, *i.e.*, the bulk sensitivity is 437.2 nm/RIU. The noise level is ~ 3 pm, and the ILOD is ~ $2.1 \times 10^{-3}$ RIU by Eq (4).

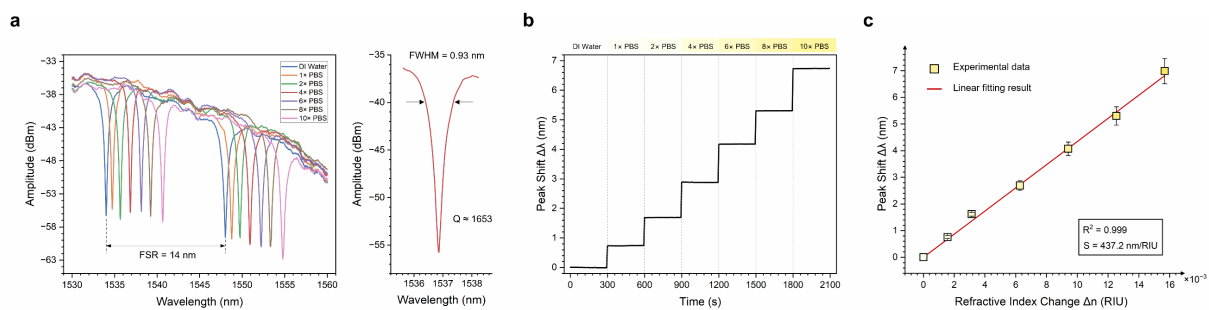

**Fig. 3. Bulk sensitivity analysis of the fabricated SUMIRRs.** The multiple represents the relative concentration compared with the standard PBS solution. **a** The transmission spectra of ring resonator in different solutions. **b** The real-time peak shift tracking with increasing solution concentration and refractive index. **c** The linear regression shows a good linear relationship between peak shift refractive index change, and the slope indicates the bulk sensitivity of device. Details about refractive indexes of PBS solution are shown in Table S4.[50]

## 4.2 Detection of SARS-CoV-2 Antigen

The viral envelope of SARS-CoV-2 consists of three structural proteins, including membrane protein (M), envelope protein (E), and spike protein (S).[60] Among these proteins, the spike protein is crucial in penetrating host cells as the major transmembrane protein.[60,61] Besides, the spike protein exhibits diversity and specificity among coronaviruses, contributing to the most immune recognition in the human body.[62] Therefore, the S protein represents an ideal target for the specific detection of SARS-CoV-2. In this study, we utilized the specific combination of S protein and SARS-CoV-2 spike antibody to realize quantitative detection. The antibody we used shows cross-reactivities with most of sub-variants of SARS-CoV-2, including Delta and Omicron.[51] After introducing samples, antibodies could combine with S proteins and form antigen-antibody complex leading to peak shifts.

We developed a protocol for surface functionalization for the diagnostic assay on the SUMIRR-based sensor platform (Fig. 4.a). The shift of resonate peak was monitored in real time throughout the above process and used to represent the extent of reaction. As shown in Fig. 4.b, the trend of peak shift in each step is similar, *i.e.*, shift increases rapidly at the beginning and then slows down until becomes flat. To further confirm that the SUMIRR surface was functionalized, we immobilized the SARS-CoV-2 spike antibody with green fluorescence and took fluorescence images for characterization. Fig. 4.c indicates that antibodies with green fluorescent coat uniformly on the waveguide, and the SWG structure manifests higher fluorescence intensity due to a larger surface area.

To investigate the performance of the SUMIRR-based biosensing platform for COVID-19 detection, we first evaluated the dynamic response of the sensor to SARS-CoV-2 spike protein with different concentrations (Fig. 5). Because we only detected one antigen in this section, the left and right sensing groups had the same function. Therefore, the reagents for surface functionalization were introduced from the common port, divided into two tributaries by splitter, and flowed to two parallel sensing groups, as shown in Fig. S8.a.

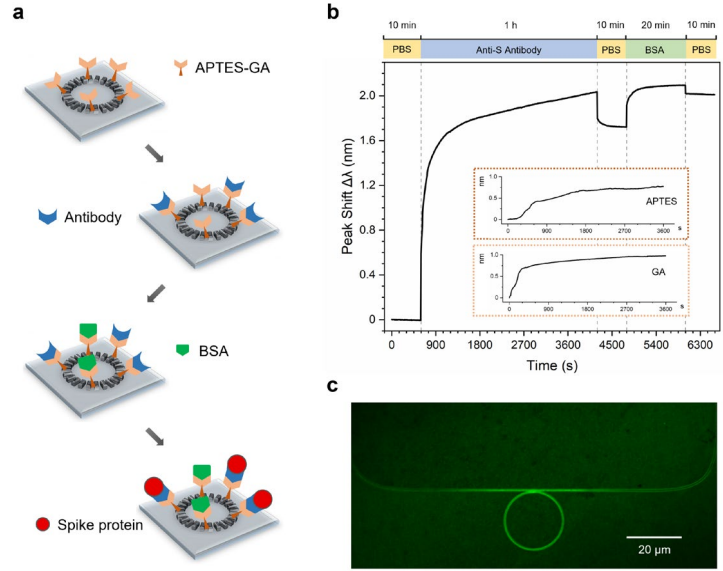

**Fig. 4. The surface functionalization of SUMIRRs. a** Schematic illustrating for surface functionalization and SARS-CoV-2 spike protein detection. **b** Real-time peak shift tracking for the surface functionalization workflow. Since the solvents for APTES and GA are ethanol and DI water, respectively, instead of standard PBS solution for the following reagents, the real-time tracking for APTES/GA coating is demonstrated individually in the inserts with different baselines. **c** Fluorescence image of the SUMIRR modified by SARS-CoV-2 spike antibody with green fluorescent.

We prepared samples with various concentrations of SARS-CoV-2 spike protein (from 10 pg/mL to 1 μg/mL) and introduced samples successively in order of concentration from low to high at 5 μL/min. The response to different samples was monitored in real time, and the dynamic tracking of peak shift was shown in Fig. 5.a. The stepwise change indicates that the peak shift within the same reaction time shows a positive correlation with antigen concentration. Besides, the negative control group without antibody coating only shows a slight response (grey line in Fig. 5.a), indicating that the reason for shift is antigen-antigen combination rather than the refractive index change due to increasing concentration. In addition, the peak shift increases rapidly at the beginning of reaction, and the increase slows down gradually. This general trend, also seen in surface functionalization, manifests that the peak shifts over a certain period of time can characterize the extent of antigen-antibody combination, thereby realizing the quantitative detection of the antigen of interest.

After demonstrating the sensing performance of the platform, we further examined the quantitative relationship between peak shift and antigen concentration for clinical detection and explored the LOD of the device. To this end, we prepared samples containing SARS-CoV-2 spike protein ranging from 100 fg/mL to 1 μg/mL. In practical detections, one sensing group can test only one antigen in the sample because the antigen cannot be removed completely once attached to the antibody. Hence, samples with different concentrations were introduced to different sensing groups instead of successively introduced to one group. The reaction time and flow rate of each sample is 10 min with a flow rate of 5 μL/min. In order to remove the antigen that was not bound to antibody and keep the refractive index of aqueous cladding layer consistent with that before the test, we used PBS solution to wash sensing surface for 5 min at 10 μL/min. The real-time response of the SUMIRR-based biosensor to a specific concentration of SARS-CoV-2 antigen is shown in Fig. 5.b. Additionally, we analyzed the peak shift for each concentration before and after washing (Fig. 5.c). The concentration-dependent response after washing fits well with the Hill model, as shown in Fig. 5.d.[63,64] These results further indicate the positive correlation between peak shift and antigen concentration, and the antigen concentration ≥ 100fg/mL could bring a remarkable response much higher than noise level.

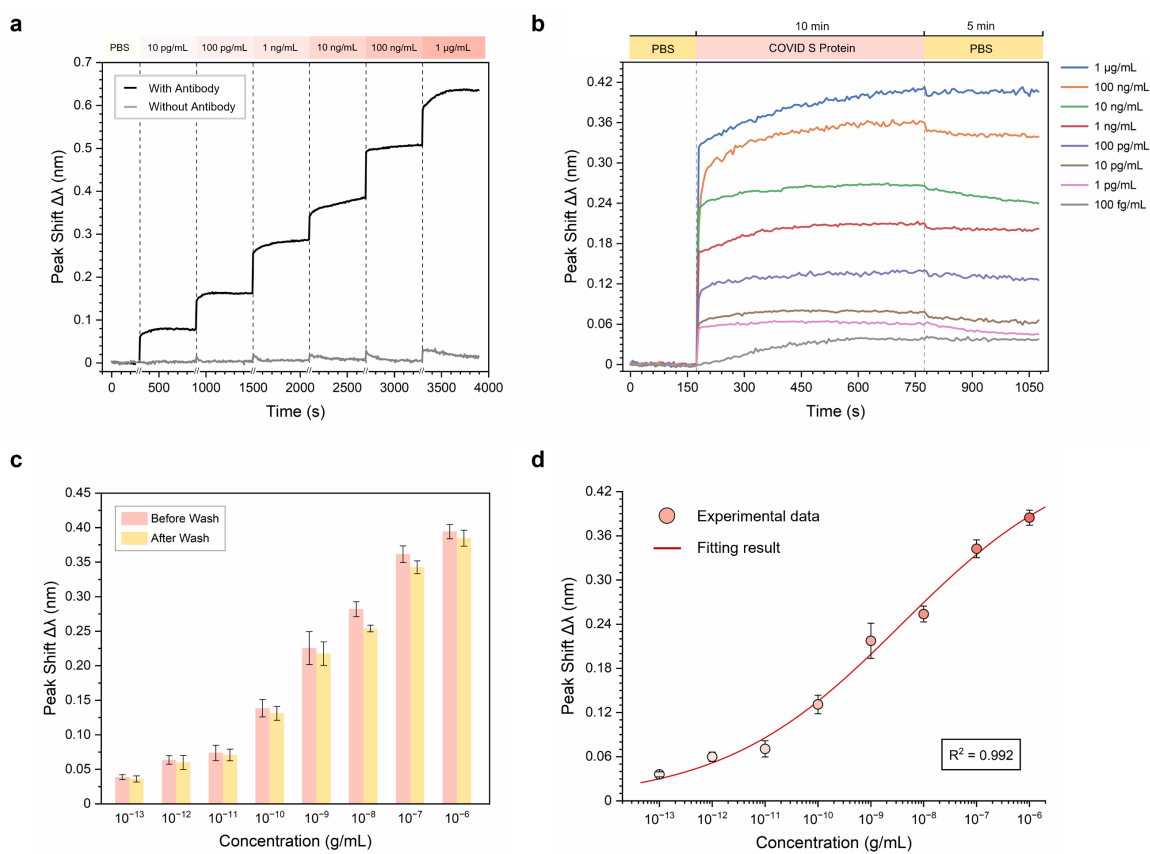

**Fig. 5. Real-time detection of SARS-CoV-2 antigen. a** Real-time response of the SUMIRR-based biosensor to SARS-CoV-2 spike protein was introduced successively by increasing order of concentration. The grey line represents the negative control group without anti-S protein antibody coating. The interval for sample replacement and PBS washing is not shown in the figure, so there are step changes. **b** Real-time response of the SUMIRR biosensor to a specific concentration of SARS-CoV-2 spike protein. **c** Resonate peak shifts due to spike protein with different concentrations before and after PBS washing. **d** Concentration-dependent response curve based on the peak shifts after PBS washing (details of the curve fitting are in the SI).

### 4.3 Specificity Analysis and Concurrent Detection

This study aimed to realize the concurrent detedction of SARS-CoV-2 and influenza viurs, so two sensing groups need to be functionalized separately. Specifically, the SARS-CoV-2 spike antibody and pan influenza A nucleoprotein antibody were introduced from the left port and right port surface for functionalization,[65] respectively, while the common port worked as the inlet during testing (Fig. S8.b). Before demonstrating the concurrent detection, we first examined the specificity of the optical assay based on antibody-antigen combination. Samples containing influenza nucleoprotein and SARS-CoV-2 spike protein were successively introduced into the sensing group functionalized by SARS-CoV-2 spike antibody, and the verification for influenza nucleoprotein antibody was taken with the same logic. The flow rate and reaction time for each step were consistent with those in the previous section. Besides, according to the antigen concentration in nasopharyngeal swab specimens from COVID-19 patients, we selected 100 pg/mL as the antigen concentration for this experiment.[23] As shown in Fig. 6.a & b, the optical assay shows a clear signal difference between positive and negative. However, the antigen that does not match the antibody also causes a slight shift (~ 15 pm), which may be related to nonspecific interaction or incomplete BSA blocking. This undesired response requires more careful determination of the LOD of sensing system. The concurrent detection results further indicate the specificity of antibody-antigen combination (Fig. 6.c). Besides, SARS-CoV-2 spike protein samples with concentrations above 100 fg/mL can result in significant peak shifts, which are distinct from nonspecific responses. Therefore, taking 100 fg/mL as the LOD of system is conservative and reliable.

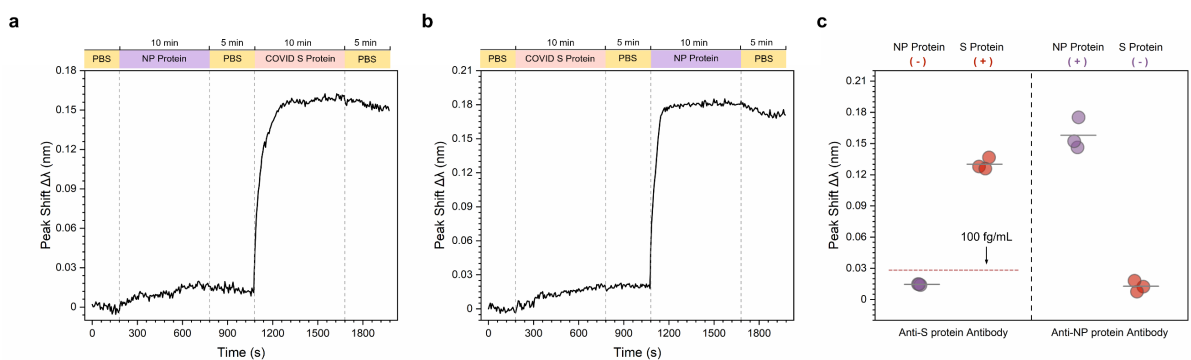

**Fig. 6.** Concurrent detection for SARS-CoV-2 spike protein and influenza nucleoprotein. **a** & **b** Specific response of SUMIRR biosensor to different antigens. The real-time responses shown in a and b were collected from the sensing groups functionalized by SARS-CoV-2 and influenza antibody, respectively. **c** Cross-reactivity tests for SARS-CoV-2 and influenza. Samples that contain SARS-CoV-2 spike protein/ influenza nucleoprotein test positive for the corresponding antibody while negative for the other antibody.

## 5. Conclusions

We have developed a POC biosensor that supports the concurrent detection and differentiation of two analytes using SUMIRRs. A microfluidic chip and an advanced packaging are developed to integrate the sensing unit, which is more convenient for clinical tests and improves reliability. The antibody immobilized on SUMIRRs could bring resonate peak shifts after combining to the antigen in sample. By analyzing the redshifts, we demonstrated the ability to quantitatively detect the concentration of SARS-CoV-2 spike protein with a LOD of 100 fg/mL. Furthermore, the cross-validation of SARS-CoV-2/influenza antigen and corresponding antibody indicates the high specificity of the optical assay based on antibody-antigen combination. Therefore, the SUMIRR-based LOC biosensor for the quantitative detection of COVID-19 provides a promising solution to overcome challenges in the rapid identification of pathogenic virus with similar symptoms, and promotes the development of POC diagnostic tools.

## Conflict of interest

The authors declare that they have no competing financial interest or associative interest that represents a conflict of interest in connection with the work reported.

## CRediT authorship contribution statement

**Shupeng Ning:** Conceptualization, Methodology, Formal analysis, Investigation, Writing Original Draft, Visualization **Hao-Chen Chang:** Methodology, Formal analysis, Supervision, Resource **Kang-Chieh Fan:** Software **Po-yu Hsiao:** Methodology, Investigation **Chenghao Feng:** Methodology, Investigation, Validation **Devan Shoemaker:** Methodology **Ray T. Chen:** Conceptualization, Writing - Review & Editing, Supervision, Project administration, Funding acquisition

## Acknowledgments

This research program is supported by Omega Optics, NIH, AFOSR MURI and NSF and the Texas State Endowment from the University of Texas, Austin.

## Appendix A. Supplementary Information